\documentclass{emulateapj}
\usepackage{graphicx}
\usepackage{amssymb}
\usepackage{epstopdf}

\long\def\symbolfootnote[#1]#2{\begingroup%
\def\thefootnote{\fnsymbol{footnote}}\footnote[#1]{#2}\endgroup}

\begin{document}

\title{Dark Stars: a new look at the First Stars in the Universe}
\author{
Douglas Spolyar\altaffilmark{1},
Peter Bodenheimer\altaffilmark{2},
Katherine Freese\altaffilmark{3},
and
Paolo Gondolo\altaffilmark{4}}
\email{ktfreese@umich.edu}
\email{peter@ucolick.org}
\email{dspolyar@physics.ucsc.edu}
\email{paolo@physics.utah.edu}

\altaffiltext{1}{Physics Dept., University of California, Santa Cruz, CA 95064}
\altaffiltext{2}{UCO/Lick Observatory and Dept. of Astronomy and Astrophysics,
 University of California, Santa Cruz, CA 95064}
\altaffiltext{3}{Michigan Center for Theoretical Physics, Physics Dept.,
Univ. of Michigan, Ann Arbor, MI 48109}
\altaffiltext{4}{Physics Dept., University of Utah, Salt Lake City, UT 84112}

\begin{abstract}
\noindent

We have proposed that the first
 phase of stellar evolution in the history of the Universe may be Dark Stars (DS),
 powered by dark matter  heating rather than by nuclear fusion, and in this paper  we 
examine the history of these DS. The power source  is annihilation of Weakly 
Interacting Massive Particles (WIMPs) which are their own antiparticles.  These WIMPs
 are the best motivated dark matter (DM) candidates and may be discovered by ongoing direct or indirect
detection searches (e.g. FERMI/GLAST) or at the Large
 Hadron Collider at CERN.  A new stellar phase results, powered by DM annihilation
 as long as there is DM fuel, from millions to billions of years. We build up the dark 
stars from the time DM heating becomes the dominant power source, accreting more and
 more matter onto them.  We have included many new effects in the current study,
 including a variety of  particle masses and  accretion rates,  nuclear burning,
 feedback mechanisms, and possible repopulation of DM density due to capture. 
 Remarkably, we find that in all these cases, we obtain the same result:  the first stars
 are very large, 500-1000 times as massive as the Sun; as well as
puffy (radii 1-10 A.U.),  bright ($10^6-10^7 L_\odot$), and cool ($T_{surf}
 < $10,000 K) during the accretion.  These results differ markedly from the standard
 picture in the absence of DM heating, in which the maximum mass is about 140
 $M_\odot$ (McKee \& Tan 2008) and the temperatures are much hotter ($T_{surf} > $50,000K).
 Hence DS should be observationally distinct from standard Pop III stars.
In addition, DS avoid the (unobserved) element enrichment produced by the standard first stars.
Once the dark matter fuel is exhausted, the DS becomes a heavy main sequence star;
these stars eventually collapse to form massive black holes that
 may provide seeds for the supermassive black holes observed at
 early times as well as explanations for recent ARCADE data (Seiffert et al. 2009) and for intermediate mass black holes.  

\end{abstract}
\keywords{Dark Matter, Star Formation, Accretion}

\section{Introduction}
Spolyar et al. (2008; hereafter Paper I) proposed a new phase
of stellar evolution: the first stars to form in the Universe may be Dark Stars, powered
by dark matter heating rather than by nuclear fusion.  Here dark matter (DM), while
constituting a negligible fraction of the star's mass, provides the energy source
that powers the star.  The first stars in the Universe  
 mark the end of the cosmic dark ages, provide the enriched gas required for later
stellar generations, contribute to reionization, and may be precursors to
black holes that coalesce and power bright early quasars.  One of the outstanding
problems in astrophysics is to investigate the mass and properties of these first stars.
 Depending on their initial masses, the stars lead very different lives, produce very different elemental enrichment,
and end up as very different objects.   In this paper,
we find the mass, luminosity, temperature, and radius of dark stars as a function 
of time: our results
differ in important ways from the standard picture of first stars without DM heating.

The DM particles considered here
are Weakly Interacting Massive Particles (WIMPS), such as the lightest supersymmetric particles or Kaluza-Klein dark matter
(for reviews, see Jungman et al. 1996; Bertone et al. 2005; Hooper  \& Profumo
 2007). 
The search for these particles is one of the major motivations for the Large Hadron
 Collider at CERN.  The particles in consideration are
their own antiparticles; thus, they annihilate among themselves in the early universe and
naturally provide the correct relic density today to explain the dark matter of the universe.
The first stars are particularly good sites for DM annihilation
because they form in  a very high density DM environment:  they form at early times
at high redshifts (density scales as $(1+z)^3$) and
in the high density centers of DM haloes.  The first stars form at redshifts $z \sim 10-50$ 
in DM halos of $10^6 M_\odot$ (for reviews see e.g. 
Ripamonti \& Abel 2005; Barkana \& Loeb 2001; Yoshida et al. 2003;
Bromm \& Larson 2004; see also
Yoshida et al. 2006) One star is thought to form inside one such
DM halo. These halos consist of 85\% DM and 15\% baryons in the form of
metal-free gas made of H and He.     Theoretical calculations
indicate that the baryonic matter cools and collapses via H$_2$     
cooling (Peebles \& Dicke 1968; Matsuda et al. 1971; 
Hollenbach \& McKee 1979) into a single small protostar 
(Omukai \& Nishi 1998) at the center of the halo.

 As our canonical values,  we use the standard  $\langle
\sigma v \rangle = 3 \times 10^{-26}\,{\rm cm^3/s}$ for the
annihilation cross section and $m_\chi = 100\,{\rm GeV}$ for the WIMP
particle mass.  In this paper we examine a variety of possible WIMP masses and cross sections,
and find that our main result holds independent of these properties. We consider 
 $m_\chi$ = 1 GeV, 10 GeV, 100 GeV, 1 TeV, and 10 TeV. This variety of WIMP masses
 may be traded for a variety of 
values of the annihilation cross section, since DM heating scales as
$\langle \sigma v \rangle /m_\chi$. 

Paper I  found that
DM annihilation provides a powerful heat source in the first stars, a
source so intense that its heating overwhelms all cooling mechanisms;
subsequent work has found that the heating can dominate over fusion as well
once it becomes important at later stages (see below). 
Paper I outlined the three key ingredients for Dark Stars:
1) high dark matter densities, 2) the annihilation products get
trapped  inside the star, and 3) DM heating wins over other cooling
or heating mechanisms.  These same ingredients are required throughout
the evolution of the dark stars, whether during the protostellar
phase or during the main sequence phase.

WIMP annihilation produces energy at a rate per unit volume 
\begin{equation}
\hat Q_{DM} = \langle \sigma v \rangle \rho_\chi^2/m_\chi ,
\label{eq:e1}
\end{equation}
where $\rho_\chi$ is the mass density of the WIMPs.
Once the gas density
of the collapsing protostar exceeds a critical value, 
most of the annihilation energy is trapped in the star.  For a 100
GeV particle,  when the hydrogen density reaches $\sim 10^{13}\, {\rm
cm}^{-3}$, 
typically 1/3 of the energy is lost to neutrinos that escape the star,
while the other 2/3 of the energy is trapped inside the star.
Hence the luminosity from the DM heating is
\begin{equation}
\label{DMheating}
L_{DM} \sim {2 \over 3} \int \hat Q_{DM} dV 
\end{equation}
where $dV$ is the volume element.   From this point forwards, the DM heating
beats any cooling mechanisms.                              
This point is the beginning of the life of the dark star, a DM powered star
which lasts until the DM fuel runs out.  The DM constitutes a negligible
fraction of the mass of the dark star (less than one percent of the total stellar mass) and
yet powers the star.  This "power of darkness" is due to the 67\% efficiency of
converting the DM mass into an energy source for the star, in contrast to the $\sim$1\% efficiency
of converting nuclear  mass into energy in the case of fusion.

In this paper we ascertain the properties of the dark star as it grows from an
initially small mass at its inception to its final fate as a very large star.
 At the beginning, when the DM heating first dominates inside the star, the dark
 star mass for our canonical case of
100 GeV mass particles is $0.6 M_\odot$ with a radius of 17 AU 
(Abel et al. 2002; Gao et al. 2007).  
Then further matter can rain down upon the star, causing the dark star to grow.
By the time the star reaches about 3 M$_\odot$, with a radius of 1--10 AU, 
 it can be approximated as an
object in thermal and hydrostatic equilibrium, with dark matter heating
balancing the radiated luminosity.
There is $\sim 1000 M_\odot$ of unstable gas ($\sim$ Jeans mass) which can
in principle accrete onto the star, unless there is some feedback to stop it. 
We build up the dark star from a few solar masses by accretion, in steps of one solar
 mass. At each stage we find the stellar structure as a polytrope
in hydrostatic and thermal equilibrium.  We continue this process of building up the DS
until the DM runs out.

As seen in equation (1), a  key element in the annihilation is knowledge of the DM
 density inside the DS.
The initial DM density profile in the $10^6 M_\odot$ halo is not enough to drive
 DM heating.
We consider three mechanisms for enhancing the density. First, as originally proposed in
 Paper 1 and confirmed in Freese et al (2008b), adiabatic contraction (AC) drives up the density. As the
 gas collapses into the star, DM is gravitationally pulled along with it.  Second, as
 the DS accretes more baryonic matter, again DM falls in along with the gas. Eventually
the supply of DM due to this continued AC runs out, and the star may heat up until
 fusion sets in.  However, there is the possibility of repopulating the DM in the DS
 due to a third source of DM, capture from the ambient medium (Iocco 2008; Freese et al. 2008c). 
 Capture only becomes
 important once the DS is already large (hundreds of solar masses), and only with the
 additional particle physics ingredient of a significant WIMP/nucleon elastic scattering
cross section at or near the current experimental bounds.  Capture can be important as
 long as the DS resides in a sufficiently high density environment. In this case the DS
lives until it leaves the womb at the untouched center of a $10^6 M_\odot$ DM halo. 
Unfortunately for the DS, such a comfortable home is not likely to persist for more
 than 10-100 Myr (if the simulations are right), though in the extreme case one may hope
 that they survive to the present day.  In this paper we consider all three
 mechanisms for DM enhancement, including DM capture.
 We remark at the outset that AC is absolutely essential to the existence of the DS.
 None of these three mechanisms would operate were it not for some AC driving up the 
density in the vicinity of the DS.

Previously (Freese et al. 2008a), we considered
one specific case, that of  100 GeV particle mass and  a constant accretion rate 
of $2 \times 10^{-3} M_\odot$/yr
with no capture, assuming that the DS stellar structure can be
 described by  a series of  $n=3/2$ polytropes.
We found that it takes $\sim 0.5$ million years to build up the star to its final mass, 
which we found to be very large, $\sim 800 M_\odot$.  While the accretion was in
 process, the
dark stars were quite bright and cool, with luminosities $L \sim 10^6 L_\odot$ and 
surface temperatures $T_{surf} < 10,000K$. These results are in contrast to the 
standard properties of the Population III
stars (the first stars).  Standard Pop III stars are predicted to form by accretion
 onto a much smaller 
protostar ($10^{-3} M_\odot$; Omukai \& Nishi 1998), to reach much 
 hotter  surface temperatures $T_{\rm surf} \gg $50,000 K during accretion,  and   at the end
of accretion  to be far less massive
($\sim 100-200 M_\odot$) due to a variety of feedback mechanisms that stop accretion of the gas
(McKee and Tan 2008). 
These differences in properties between dark stars and standard Pop III stars lead
to a universe that can look quite different, as discussed below.

In this paper we present the results of  a complete  study of building up the dark
star mass and finding the stellar structure at each step in mass accretion.
We include many new effects in this paper compared to Freese et al.\ (2008a).  (1) We generalize the particle physics
 parameters by considering a variety of particle masses or equivalently annihilation
 cross sections.  (2) We study  two    possible accretion rates: 
  a variable accretion rate from Tan \& McKee (2004) and an alternate 
variable accretion rate found by O'Shea \& Norman (2007).   (3) Whereas
 previously we considered only $n=3/2$ polytropes, which should be used only when
 convection dominates, we now allow variable polytropic index to include the transition
to  radiative transport ($n=3$).  (4) We include gravity as a heat source. In the later
 stages of accretion, as the DM begins to run out and the star begins to contract,
 the gravitational potential energy is converted to an important power source. 
(5) We include nuclear burning, which becomes important once the DM starts to run out. 
(6) We include feedback mechanisms which can prevent further accretion. Once the stellar
 surface becomes hot enough, again when the DM is running out, the radiation can prevent
 accretion.  (7) We include the effects of repopulating the DM
inside the star via capture from the ambient medium.  The result of this
 paper is that,
for all the variety of parameter choices we consider, and with all the physical effects included,
with and without capture, our basic result is the same.  The final stellar mass is 
driven to be very high and, while DM reigns, the star remains bright but cool.

Other work on dark-matter contraction, capture, and
  annihilation in the first stars
(Freese et al. 2008c; Taoso et al. 2008; Yoon et al. 2008; Iocco et al.
 2008; Ripamonti et al. 2009) has concentrated on
the case of the pre-main-sequence, main-sequence, and post-main-sequence
 evolution of stars at
fixed mass, in contrast to the present work, which allows for accretion 
from the dark-matter halo.

We also cite previous work on DM annihilation in today's stars 
(less powerful than in the first stars): Krauss et al.\ (1985);
Bouquet \& Salati (1989); Salati \& Silk (1989); Moskalenko \& Wai (2007);
Scott et al.\ (2007); Bertone \& Fairbairn (2007); Scott et al.\ (2009).

\section{Equilibrium Structure}

\subsection{Basic Equations}

At a given stage of accretion, the dark star has a prescribed mass. 
We make the assumption that it       
can be described as  a polytrope in hydrostatic  and thermal equilibrium. 
The hydrostatic equilibrium is defined by 
\begin{equation}
{dP \over dr} = - \rho {GM_r \over r^2}; ~~~{dM_r \over dr} = 4 \pi r^2 \rho(r)
\end{equation} 
 and by the polytropic assumption
\begin{equation}
\label{eq:polytrope}
P = K \rho^{1 + 1/n} 
\end{equation}
where $P$ is the pressure, $\rho$ is the density,  $M_r$ is the mass enclosed within
radius $r$, and  the constant $K$ is
determined once the total mass and radius are specified (Chandrasekhar
1939).  Pre-main-sequence stellar models are adequately described by polytropes
in the range $n=1.5$ (fully convective) to $n=3$ (fully radiative). 
Given $P$ and $\rho$ at a point, the temperature $T$ is defined by the equation
of state of a mixture of gas and radiation
\begin{equation}
\label{eq:eqnofstate}
P(r) = {{R_{\rm g}\rho (r) T(r)}\over {\mu}} + {1 \over 3} aT(r)^4 
= P_g + P_{rad}
\end{equation}
where $R_{\rm g} = k_B/ m_u$ is the gas constant, 
$m_u$ is the atomic mass unit, $k_B$ is the Boltzmann constant, 
 and the mean atomic weight $\mu    
 = (2X + 3/4 Y)^{-1} =0.588$.  We take the H mass fraction $X=0.76$
 and the He     mass fraction $Y=0.24$.  In the resulting models   
 $T\gg$10,000 K except near the very surface, 
so  the approximation for $\mu$ assumes that  H and He    are  fully ionized.
In the final models with masses near 800 M$_\odot$ the radiation pressure
is of considerable importance.

Given  a mass $M$ and an estimate for the outer radius $R_\ast$, the hydrostatic
structure is integrated outward from the center. At each point $\rho(r)$ and 
$T(r)$ are used to determine the Rosseland mean opacity $\kappa$ from a zero-metallicity
table from OPAL (Iglesias \& Rogers 1996), supplemented by a table from
Lenzuni et al. (1991) for $T < 6000$ K. The photosphere is defined by the 
hydrostatic condition
\begin{equation}
\kappa P = \frac{2}{3} g
\end{equation}
where $g$ is the acceleration of gravity. This point  
corresponds to inward integrated optical depth $\tau \approx 2/3$. When
it         is reached, the local temperature is set to $T_{\rm eff}$ and
the stellar radiated luminosity is therefore 
\begin{equation}
L_\ast = 4 \pi R_\ast^2 \sigma_B T_{\rm eff}^4  
\end{equation}
where $R_\ast$ is the photospheric radius.

The thermal equilibrium condition is 
$L_\ast = L_{{\rm tot}},$
where the total energy supply for the star (see below) is dominated during
the DS phase by the DM luminosity $L_{DM}$.
Thus the next step is to determine $L_{DM}$ from equation (\ref{DMheating})
and from the density distribution of dark matter  (discussed in the next
subsection). Depending on the value of $L_{DM}$ compared to $L_\ast$, the
radius is adjusted, and the polytrope is recalculated. The dark matter
heating is also recalculated based on the revised baryon density distribution. 
The radius is iterated until the condition of thermal equilibrium is met.

\subsection{DM Densities}

The value of the DM density inside the star is a key ingredient in DM heating.
We start with a $10^6$  M$_\odot$ halo 
composed of 85\% DM     and 15\% baryons.
We take an initial Navarro, Frenk, \& White profile (1996; NFW)
with a concentration parameter $c=2$ at $z=20$
in a standard $\Lambda$CDM universe.
We note at the outset that our results are {\it not} dependent on high central DM densities
of NFW haloes.  Even if we take the initial central density to be a core (not rising towards
the center of the halo), we obtain qualitatively the same results of this paper;
we showed this in Freese et al (2008b).

{\it Enhanced DM density due to adiabatic contraction:}
Further density enhancement is required inside the star in order for annihilation to play any role.
Paper I recognized a key effect that increases the DM density: adiabatic contraction (AC).  
As the gas falls into the star, the DM is gravitationally pulled along with it.   Given the initial NFW 
profile, we follow its response to the changing baryonic gravitational potential as the
 gas condenses.  Paper I used a simple Blumenthal method, which assumes circular
 particle orbits (Blumenthal et al. 1986; Barnes \& White 1984; Ryden \& Gunn 1987)
 to obtain estimates of the density. Subsequently Freese  et al. (2008b) did an exact
 calculation using the Young method (Young 1980) which includes radial orbits, 
 and confirmed our original results (within a factor of two).  Thus we feel confident that
 we may use the simple Blumenthal method in our work.  We found
\begin{equation}
\label{eq:AC}
\rho_\chi \sim 5 {\rm (GeV/cm^3)} (n_h/{\rm cm}^3)^{0.81} ,
\end{equation}
where $n_h$ is the gas density. For example, due to this contraction,
at a hydrogen density of $10^{13}$/cm$^3$, the DM density is $10^{11}$ GeV/cm$^3$.
Without adiabatic contraction, DM heating in the first stars would be so small as to be irrelevant.  

{\it Enhanced DM due to accretion:}
As further gas accretes onto the DS, driving its mass up from $\sim 1 M_\odot$
to many hundreds of solar masses, more DM is pulled along with it into the star. 
At each step in the accretion process, we compute the resultant DM profile
in the dark star by using the Blumenthal et al. (1986)
prescription for adiabatic contraction.
The DM density profile is calculated at each iteration 
of the stellar structure, so that the DM luminosity can be determined. 

The continued accretion of DM sustains the dark star for $\sim 10^6$ years.  Then,
 unless the DM is supplemented further, it runs out, that is, it annihilates faster
than it can be resupplied.
At this point the $10^6 M_\odot$ DM halo has a tiny ``hole'' in the middle that is devoid of DM.
One might hope to repopulate the DM by refilling the ``hole,''
but this requires further study. On a conservative note, this paper disregards this possibility.
Without any further DM coming into the star
(which is now very large), it would contract and heat up to the point where fusion 
sets in; then the star joins the Main Sequence (MS).  This is one possible scenario which
 we examine in the paper.
Alternatively, there is another possible mechanism to replenish the DM in the dark star
 that may keep it living longer: capture.  

{\it Enhanced DM due to capture:}
At the later stages, the DM inside the DS may be repopulated due to capture (Iocco 2008;
Freese et al. 2008c).
This mechanism relies on an additional piece of particle
physics, scattering of DM particles off the nuclei in the star. Some of the DM particles
bound to the $10^6 M_\odot$ DM halo occasionally pass through the DS, and if they lose
 enough energy by scattering, can be trapped in the star.  Then more DM would reside
 inside the DS and could continue to provide a heat source.  This capture process is irrelevant
 during the initial stages of the DS, and only becomes important once the gas density is 
high enough to provide a significant scattering rate, which happens when the DS is
 already quite large ($> 100 M_\odot$).   For capture to be significant, there are two
 assumptions: (1) the scattering cross section must be high enough and (2) the ambient 
density of DM  must be high enough.
Regarding the first assumption:
Whereas the existence of DS relies only on a standard annihilation cross section (fixed by 
the DM relic density),
the scattering cross section is more speculative.  It can vary over many orders of
 magnitude, and is constrained only by experimental bounds from direct detection
 experiments (numbers discussed below).
Regarding the second assumption:
The ambient density is unknown, but we can again obtain estimates using the adiabatic
 contraction method of Paper 1. Indeed it seems reasonable for the ambient DM density
 to be high
 for timescales up to at most tens-hundreds of
 millions of years, but after that the $10^6 M_\odot$ host for the dark star surely has
 merged with other objects and the central regions are likely to be disrupted; simulations
 have not yet resolved this question.  Then it becomes more difficult for the DS to remain in a high
 DM density feeding ground and it may run out of fuel. 

We therefore consider two different situations:  adiabatic contraction
with and without capture; i.e., with and without additional capture of DM at the
later stages. In the latter case
we focus on ``minimal capture,'' where the stellar luminosity has equal contributions
from DM heating and from fusion, as described further below.  In both cases,
we find the same result:  the dark stars continue to accrete until they are very heavy.

\subsection{Evolution}
\label{sect:evo}
The initial condition for our simulations is a DS of 3 M$_\odot$, roughly the
mass where the assumption of nearly complete ionization first holds.  We find an equilibrium
solution for this mass, then build up the mass of the star in increments of one solar mass.

We allow surrounding matter from the original baryonic core
 to accrete onto the DS, with two different assumptions for the mass 
accretion:
first, the variable rate from Tan \& McKee (2004) and second,
the variable rate from  O'Shea \& Norman (2007). The Tan/McKee rate decreases 
from $1.5 \times 10^{-2}$ M$_\odot/{\rm yr}$ at  a DS mass of 3 M$_\odot$ to 
$1.5 \times 10^{-3}$ M$_\odot/{\rm yr}$ at   1000 M$_\odot$. The O'Shea/Norman
rate decreases from $3 \times 10^{-2}$ M$_\odot/{\rm yr}$ at  a DS mass of 3 M$_\odot$
 to $3.3 \times 10^{-4}$ M$_\odot/{\rm yr}$ at   1000 M$_\odot$. The accretion
luminosity arising from infall of material from the primordial core onto the DS
is not included in the radiated energy. It is implicitly assumed to be radiated
away in material outside the star, for example an accretion disk (McKee \& Tan
2008), and in any case,
at the large radii involved here, this energy is small compared with the DM
annihilation energy.

 We remove the amount of DM that has annihilated at each stage at each
radius.  We continue  stepping up in mass, in increments
 of 1 M$_\odot$, until we reach 1000 M$_\odot$,
the Jeans mass of the core (Bromm \& Larson 2004) or until all of the following
conditions are met: 1) the total mass is less than 1000 M$_\odot$, 2) 
the dark matter available from adiabatic contraction is used up; 
3) nuclear burning has set in, and 4) the energy generation from 
gravitational contraction is less than 1\% of the total luminosity of the
star. The last two of these conditions define the zero-age main sequence (ZAMS).
 The time step is adjusted for
each model to give the correct assumed accretion rate.
The final mass can be less than 1000 M$_\odot$ because of feedback effects
of the ionizing stellar radiation, which can reverse the infall of
accreting gas (McKee \& Tan 2008). The feedback effect is included in an
approximate manner, starting when the surface temperature reaches 50,000 K.

 At the beginning of the evolution, the Schwarzschild stability criterion for
convection shows that the model is fully convective; thus the polytrope of
$n=1.5$ is a good approximation to the structure. However as more mass is added
and internal temperatures increase, opacities drop and, once the model has
reached about 100 M$_\odot$, radiative zones start to appear at the inner radii and
move outward. Beyond about 
300 M$_\odot$ the model is almost fully radiative. The evolutionary calculation
takes into account this effect approximately,  
by gradually shifting the value of $n$ up from
1.5 to 3 in the appropriate mass range. In the later stages of evolution, the
constant value $n=3$ is used.

\subsection{Energy Supply}
The energy supply for the star changes with time and comes from four major
sources: 
\begin{equation}
L_{\rm tot}=L_{DM} + L_{\rm grav} + L_{\rm nuc} + L_{\rm cap} .
\end{equation}
 The general
thermal equilibrium condition is then 
\begin{equation}
L_\ast = L_{\rm tot} . 
\end{equation}
The  contribution $L_{DM}$ from adiabatically contracted DM dominates the heat supply first.
Its contribution been discussed previously and is given in equation (\ref{DMheating}).
Later, once the adiabatically contracted DM becomes sparse, the other effects kick in:
gravity, nuclear burning, and captured DM.
We discuss each of these contributions in turn.
  
 \subsubsection{Gravity}
 At the later stages of evolution, when the supply of DM starts to run out,
the star adjusts to the reduction in heat supply by contracting. For a
polytrope of $n=3$, the total energy of the star is
\begin{equation}
E_{\rm tot} = {-} \frac{3}{2} \frac{GM^2}{R} + E_{\rm th} + E_{\rm rad}
\label{etot}
\end{equation}
where $E_{\rm th}$ is the total thermal energy and  the radiation energy
$E_{\rm rad} = \int aT^4 dV$. 
Using the virial theorem
\begin{equation}
 {-} \frac{3}{2} \frac{GM^2}{R} + 2 E_{\rm th} + E_{\rm rad} = 0 ,
\end{equation}
we eliminate $E_{\rm th}$ from equation (\ref{etot}) and obtain the
gravitational contribution to the radiated luminosity
\begin{equation}
 L_{\rm grav} = {-} \frac{d}{dt}(E_{\rm tot}) =  \frac{3}{4} 
\frac{d}{dt}\left(\frac{GM^2}{R}\right) - \frac{1}{2} \frac{d}{dt} E_{\rm rad} \, .
\end{equation}

\subsubsection{Nuclear Fusion}
Once the star has contracted to sufficiently high internal temperatures $\sim 10^8$K,
nuclear burning sets in. For a metal-free star, we include the following
three energy sources: 1) the burning of the primordial
deuterium (mass fraction $4 \times 10^{-5}$) at temperatures around $10^6$ K,
2) the  equilibrium proton-proton cycle for hydrogen burning, and 3) the triple-alpha reaction
for helium burning.  At the high internal temperatures ($2.75 \times 10^8$ K)
needed for the star to reach the ZAMS, helium burning is an important
contributor to the energy generation, along with the proton-proton cycle.
We stop the evolution of the star when it reaches the ZAMS, so changes in abundance of H and He
are not considered.
Then $L_{\rm nuc} = \int \epsilon_{\rm nuc} dM$ where $\epsilon_{\rm nuc}$ is
the energy generation rate in erg g$^{-1}$ s$^{-1}$; it is obtained by the
standard methods described in Clayton (1968). The astrophysical cross
section factors for the proton-proton reactions are taken from 
Bahcall (1989, Table 3.2), and the helium-burning parameters are taken
from Kippenhahn \& Weigert (1990).

\subsubsection{Feedback}
During the initial annihilation phase, the surface temperature $T_{\rm eff}$ remains at or below 
$10^4$ K. However once gravitational contraction sets in, it increases
substantially, reaching $\approx 10^5$ K when nuclear burning starts.
In this temperature range, feedback effects from the stellar radiation
acting on the infalling material can shut off accretion (McKee \& Tan 2008).
These effects include photodissociation of H$_2$, Lyman $\alpha$ radiation
pressure, formation and expansion of an HII region, and photoevaporation of
a disk. These effects become important at a mass of $\approx 50$ M$_\odot$
in standard Pop III calculations of accretion and evolution, but the
effect is suppressed until much higher masses in the DM case because of
much lower $T_{\rm eff}$. Noting that 50  M$_\odot$ on the standard
metal-free ZAMS      corresponds to $T_{\rm eff} \approx $ 50,000 K
(Schaerer 2002), we apply a linear reduction factor to the accretion rate above
that temperature, 
such that accretion is shut off completely when $T_{\rm eff} =$100,000 K. 
This   cutoff generally determines the final mass of the star when it reaches 
the ZAMS. 

\subsubsection{Capture}
The contribution $L_{\rm cap}$ is determined by calculating the annihilation rate of captured DM
\begin{equation}
L_{\rm cap}=2 m_\chi \Gamma_{\rm cap} ,
\end{equation}
where 
\begin{equation}
\Gamma_{cap}=\int d^3x \rho_{\rm cap}^2 \langle \sigma v(r) \rangle /m_\chi
\end{equation}
 and
$\rho_{\rm cap}(r)$ is the  density profile of captured DM.  DM heating from captured DM has 
previously been discussed in detail by Freese et al. (2008c) and Iocco (2008).  

The density profile of the captured DM  has  quite a different shape
 than the adiabatically contracted (AC) DM density profile discussed previously.
 Whereas the AC profile is more or less constant throughout most of the star (see Fig. 3),
the captured DM profile is highly concentrated toward the center with an exponential
falloff in radius, as we will now show.
 Once  DM is captured, it scatters multiple 
 times, thermalizes with the star,
 and sinks to the center of the star. 
The DM  then assumes the form 
 of a  thermal distribution in the gravitational well of the star  
\begin{equation}
\rho_{\rm cap}=m_\chi n_{\rm cap}=f_{\rm Eq} f_{\rm Th}\rho_{\rm o, c}e^{-[\phi(r) m_\chi/ T_c]} \, .
\label{eq:rhocap}
\end{equation}
Here $T_c$, the star's central temperature, is used since the DM has the same temperature as the stellar core;
$\phi(r)$ is the star's gravitational potential;
 $\rho_{\rm o,c}$ is the central density of the captured DM; and $f_{\rm Th}$ and $f_{\rm Eq}$  are discussed below.  One can immediately see another difference with the AC profile:  for captured DM the
 profile is more centrally concentrated for larger DM particle mass, whereas the density profile for adiabatically contracted DM is independent of particle mass. 

The quantity $f_{\rm Th}$  is the probability  that a captured DM particle has 
thermalized with the star. 
\begin{equation}
f_{\rm Th}=1-e^{-t_1/\tau_{\rm Th}}.
\end{equation}
$t_1$ is  the time since a DM particle has been captured
 and $\tau_{\rm Th} $ is the thermalization time scale of captured DM.    

 Captured DM does not
 immediately thermalize with the star; it requires multiple scatters.  
 The thermalization time scale for the DM is
\begin{equation}
\tau_{\rm Th}= \frac{m_\chi/m_{\rm H}}{2 \sigma_{\rm sc} v_{\rm esc} n_{H}}
\end{equation}
as previously discussed in Freese et al. (2008c).  Here
 $v_{\rm esc}$ is the escape 
velocity from the surface of the star, $n_{H}$ is 
the average stellar density of hydrogen, and $m_{\rm H}$ is the proton mass.
 We have only considered
spin-dependent scattering, which includes scattering off hydrogen
 with spin 1/2, but not off Helium which has spin 0.

 Heating from captured DM 
is unimportant until the star approaches the MS; the 
thermalization time scale is very long  (millions of years) and $f_{\rm 
Th}\approx0$.
  Once the DM supplied from adiabatic contraction runs out,
   the star begins to contract, and the thermalization time scale becomes 
very short, on the order of a year.
Thus, the star rapidly transitions from $f_{\rm Th}\approx 0$
 to $f_{\rm Th}\approx 1$.
 The DM is either in thermal equilibrium or it is not.  With this in
 mind, $t_1$ has been set to $100~{\rm yrs}$ for all of the simulations.  Only $\tau_{\rm Th}$ 
 is recalculated at every time step, which is a reasonable prescription;
  $t_1$ set between a few thousand years and a few years makes no 
 quantitative difference upon the evolution of the DS.

The ratio $f_{\rm Eq}$ gives the ratio of DM particles  currently inside
 of the star due to capture,  $N$,  to   the maximum 
number of DM particles that could possibly be present,  $N_o$,
\begin{equation}
f_{\rm Eq}=\frac{N}{N_o}.
\end{equation}
 At the start of the simulation, the DS is too diffuse to capture any DM and $f_{\rm Eq}\approx0$. Once the star goes onto the main sequence, the 
maximum  number is reached and $f_{\rm Eq}\approx1$.
 
We can now give an explicit expression of $f_{\rm Eq}$ by solving for the total number of particles $N$
bound to the star.
 This number  is determined by a competition between
  capture and annihilation,
 \begin{equation}
 \label{eq:dotn1}
 \dot N=C-2 \Gamma \, ,
 \end{equation}
where `dot' refers to differentiation in time. 
The total number of DM particles captured 
per unit time $C$ was calculated following Gould (1987a).
 The total number of DM particles
 annihilated per unit time $\Gamma$ contains a factor of two since two 
particles are annihilated  per annihilation.). 
Equation (\ref{eq:dotn1})  can be solved for $N$,
\begin{equation}
\label{equilibrium}
N=N_o \tanh(t_2/\tau_{eq})
\end{equation}
where
$ N_o=\sqrt{C/C_A},$
and
$\tau_{eq}=1/\sqrt{ C C_A}$, and  ${C_A}=2\Gamma/{N^2}$ 
 (Griest \& Seckel 1987).

When the annihilation rate equals the capture rate,  $N$ goes to  $N_o$. This transition occurs on the equilibrium time scale $\tau_{eq}$.  The equilibrium time scale is recalculated at each time step.  
   The time $t_2$ in equation (\ref{equilibrium}) counts the time since
 capture is turned on.
  In equation (\ref{eq:rhocap}), we now set
\begin{equation}
f_{\rm Eq}=\frac{N}{N_o}=\tanh(t_2/\tau_{eq}).
\end{equation} 
   
 Conceptually $t_2$ is not the same as $t_1$ but at a practical level  it is reasonable to set $t_2=t_1$.
In the first place, $t_1$ refers to the time since a single DM particle has been captured and in fact  $t_1$ is different for 
each captured DM particle. On the other hand,  $t_2$ is a single time for all captured and thermalized DM and starts when 
capture turns on.    
 Regardless,  once the star descends towards the main sequence, $f_{\rm Eq}$ and $f_{\rm Th}$ both go to unity.  The connection is not a coincidence.  First $\tau_{eq}$ is a factor of a few to a hundred less than $\tau_{\rm Th}$ throughout the simulation.  In addition, $t_2 > t_1$ except for the very first DM particle captured.  Fixing $t_2=t_1$ does not change the behavior of $f_{\rm Eq}$ in relation $f_{\rm Th}$ as just described.
Hence, for simplicity the simulation assumes $t_1=t_2$.  

In the limit that $f_{\rm Eq}$ and $f_{\rm Th}$ go to unity, then 
  $\Gamma_{cap}=C/2$. Hence,
 $L_{\rm cap}$ simplifies to
\begin{equation}
L_{\rm cap}=\frac{2}{3} m_\chi C
\end{equation}
where the $2/3$ accounts for the loss of neutrinos
 as previously discussed in connection with equation (\ref{DMheating}).
 Also roughly
 $L_{\rm cap}\propto \bar\rho_\chi \sigma_{sc}$ and is independent of $m_\chi$.  Hence $L_{\rm cap}$ scales with both
the  ambient background DM density $\bar\rho_\chi$, and  the scattering cross section of the DM off of baryons $\sigma_{\rm sc}$.

Physically reasonable values for the scattering cross section and 
background DM densities imply that $L_{\rm cap}$  can affect the first stars once they
approach the MS.
Direct and indirect detection experiments require
$\sigma_{sc}\leq4\times 10^{-39}$ (Savage et al. 2004; Chang et al. 2008; Savage et al 2008; Hooper et al. 2009);  
a WIMP's $\sigma_{\rm sc}$ can be many orders of 
magnitude smaller than the experimental bound. 
The ambient background DM density,  $\bar\rho_\chi$, can
 be very high,  on the order of $10^{14} {\rm (GeV/cm^3)}$ (as obtained from Eq. (8))
 due to adiabatic contraction. 
 To demonstrate the ``minimal" effect of capture, we set
$\bar\rho_\chi=1.42\times10^{10} {\rm (GeV/cm^3)}$ and
 $\sigma_{sc}=10^{-39} {\rm cm^2}$, chosen so that about  half of the 
luminosity is from DM capture and the other half from fusion for the standard 100 GeV 
case at the ZAMS. These parameters have been used in all cases where
 capture has been turned on.
Even    minimal capture  shows that DM capture 
lowers the star's surface temperature and increases the mass of the first stars.

 We note that the product of ambient density and cross section could be very different
 than the values used in this paper.  If this product is much smaller than
 the values considered in the 'minimal capture' case, then fusion completely dominates
 over capture and one may ignore it.  On the other hand, if this product is much larger than
 the 'minimal capture' case, then fusion is negligible and the DS may continue to
 grow by accretion and become
 much larger than discussed in this paper; this case will be discussed elsewhere.
 The 'minimal' case is thus the borderline case between these two possibilities, and
 lasts only as long as the DS remains embedded in  a high ambient DM density
 assumed here.
 
\section{Results}

\begin{table*}
 \caption{ Properties and Evolution of Dark Stars for $m_\chi = 1$
GeV, 
$\dot M $ from Tan \& McKee, $\langle \sigma v \rangle = 3 \times 10^{-26}$ cm$^3$/s. 
For the lower stellar masses, identical results are obtained for the cases of no capture and minimal
capture; hence the first four rows apply equally to both cases.  The fifth and sixth rows
apply only to the no capture case while the last two rows apply only when capture is included.
}
\begin{center}
{
\small
 \begin{tabular}{||l|l|l|l|l|l|l|l|l|c||}
 \hline\hline
$ M_\ast$ & $L_\ast$ &  $R_\ast$ & $T_{\rm eff}$ & 
$\rho_c $ & $T_c $& $M_{DM} $ & $\rho_{\rm o,DM} $& $t$ \\ 
$(M_\odot)$ & $(10^6 {\rm L_\odot})$ & $(10^{12}{\rm cm})$ & $(10^3 {\rm K})$ & 
$({\rm gm/cm}^3)$ & $(10^6 {\rm K})$  &$(10^{31} {\rm g})$ &$({\rm gm/cm}^3)$ & ($10^3$yr)  \\ 

\hline
  106 
& $9.0$
& $240 $
& $5.4$
& $1.8 \times 10^{-8}$
& $0.1$
& $36$
&$4.5\times10^{-11}$
& $19$\\
\hline
  371
& $17$
& $270$
& $5.9$
& $5.2 \times 10^{-8}$
& $0.20$
& $64$
&$2.0\times10^{-11}$
& $119$\\
\hline
690
&$6.0$
&$110$
&$7.5$
&$5.9\times 10^{-6}$
&$1.0$
&$10$
&$6.8\times10^{-11}$
&$280$\\
\hline
  756
& $3.3$
& $37$
& $10$
& $2.2 \times 10^{-4}$
& $3.4$
& $1.2$
&$4.8\times10^{-10}$
& $310$\\
\hline
  793
& $4.5$
&$5.1$
& $31$
& $8.9\times10^{-2}$
& $25$
& $\approx0.0$
& $\approx0.0$
&$330$\\
\hline
 820
& $8.4$
& $0.55$
& $111$
& $115$
& $276$
& $\approx0.0$
& $\approx0.0$
& $430$
\\
\hline
\hline
Alternative properties \\ for final two masses\\
with capture included: \\
\hline
\hline
793
&$4.6$
&$5.7$
 &$30$
 &$7.1\times10^{-2}$
 &$24$
&$8.4\times10^{-4}$
&$2.94\times10^{-8}$
 &$328$\\
 \hline
 824
 &$8.8$
 &$.58$
 &$109$
 &$102$
 &$265$
&$8.1\times10^{-5}$
 &$2.14\times10^{-6}$
 &$459$\\
 \hline 
 \hline\hline
 \end{tabular} 
 \label{tab:ExpParam1}
}
\end{center}
\end{table*}

\begin{table*}
 \caption{ Properties and Evolution of Dark Stars for $m_\chi = 100$
GeV, 
$\dot M $ from Tan \& McKee, $\langle \sigma v \rangle = 3 \times 10^{-26}$ cm$^3$/s.  Entries
as described in Table 1.
}
\begin{center}
{
\small
 \begin{tabular}{||l|l|l|l|l|l|l|l|l|c||}
 \hline\hline
$ M_\ast$ & $L_\ast$ &  $R_\ast$& $T_{\rm eff}$ & 
$\rho_c $ & $T_c $& $M_{DM} $ & $\rho_{\rm o,DM} $& $t$ \\ 
$(M_\odot)$ & $(10^6 {\rm L_\odot})$ & $(10^{12}{\rm cm})$ & $(10^3 {\rm K})$ & 
$({\rm gm/cm}^3)$ & $(10^6 {\rm K})$  &$(10^{31} {\rm g})$ &$({\rm gm/cm}^3)$ & ($10^3$yr)  \\ \hline
  106 
& $1.0$
& $70 $
& $5.8$
& $1.1 \times 10^{-6}$
& $0.4$
& $16$
&$1.8\times10^{-9}$
& $19$\\
\hline
  479
& $4.8$
& $84$
& $7.8$
& $2.0 \times 10^{-5}$
& $1.4$
& $41$
&$2.0\times10^{-9}$
& $171$\\
\hline
600
&$5.0$
&$71$
&$8.5$
&$4.1\times10^{-5}$
&$1.8$
&$36$
&$1.9\times10^{-9}$
&$235$\\
\hline
  716
& $6.7$
& $11$
& $23$
& $2.0 \times 10^{-2}$
& $15$
& $2.7$
&$3.8\times10^{-8}$
& $303$\\
\hline
  756
& $6.9$
&$2.0$
& $56$
& $2.8$
& $78$
& $\approx0.0$
& $\approx0.0$
&$330$\\
\hline
 779
& $8.4$
& $0.55$
& $110$
& $120$
& $280$
& $\approx0.0$
& $\approx0.0$
& $387$\\
\hline
\hline
Alternative properties \\for final two masses \\
with capture included: \\
\hline
\hline
756 (c)
&$6.8$
&$2.0$
 &$55$
 &$2.55$
 &$76$
&$7.5\times10^{-5}$
&$5.3\times10^{-5}$
 &$328$\\
 \hline
 787(c)
 &$8.5$
 &$0.58$
 &$108$
 &$105$
 &$270$
&$2.2\times10^{-5}$
&$5.3\times10^{-4}$
 &$414$\\
 \hline 
 \hline\hline
 \end{tabular} 
 \label{tab:ExpParam2}
}
\end{center}
\end{table*}
%
%
%
%
\begin{table*}
 \caption{ Properties and Evolution of Dark Stars for $m_\chi = 10$
TeV, 
$\dot M $ from Tan \& McKee, $\langle \sigma v \rangle = 3 \times 10^{-26}$ cm$^3$/s. 
Entries as described in Table 1.
}
\begin{center}
{
\small
 \begin{tabular}{||l|l|l|l|l|l|l|l|l|c||}
 \hline\hline
$ M_\ast$ & $L_\ast$ &  $R_\ast$ & $T_{\rm eff}$ & 
$\rho_c $ & $T_c $& $M_{DM} $ & $\rho_{\rm o,DM} $& $t$ \\ 
$(M_\odot)$ & $(10^6 {\rm L_\odot})$ & $(10^{12}{\rm cm})$ & $(10^3 {\rm K})$ & 
$({\rm gm/cm}^3)$ & $(10^6 {\rm K})$  &$(10^{31} {\rm g})$ &$({\rm gm/cm}^3)$ & ($10^3$yr)  \\ 
\hline
  106 
& $0.1$
& $22$
& $6.0$
& $3.8 \times 10^{-5}$
& $1.3$
& $7.8$
&$1.8\times10^{-8}$
& $19$\\
\hline
  310
& $0.59$
& $21$
& $9.3$
& $4.4\times 10^{-4}$
& $3.6$
& $15$
&$1.5\times10^{-7}$
&$88$\\
\hline
399
&$3.3$
&$6.6$
&$25$
&$3.6\times10^{-2}$
&$16$
&$6.6$
&$9.9\times10^{-7}$
&$134$\\
\hline
  479
& $3.0$
& $2.9$
& $32$
& $0.5$
& $40$
& $2.1$
&$1.1\times10^{-6}$
& $172$\\
\hline
 552
& $5.2$
& $0.56$
& $97$
& $79$
& $220$
& $\approx0.0$
& $\approx0.0$
& $230$\\
\hline
 552
& $5.3$
& $0.47$
& $107$
& $136$
& $270$
& $\approx0.0$
& $\approx0.0$
& $256$
%
\\
\hline
\hline
Alternative properties \\ for final two masses  \\
with capture included: \\
\hline
\hline
550 (c)
&$5.2$
&$0.6$
 &$95$
 &$66$
 &$210$
&$8.5\times10^{-12}$
&$5.3\times10^{4}$
 &$230$\\
 \hline
 553 (c)
 &$5.4$
 &$0.48$
 &$105$
 &$127$
 &$260$
&$5.7\times10^{-12}$
&$1.2\times10^5$
 &$266$\\
 \hline 
 \hline\hline
 \end{tabular} 
 \label{tab:ExpParam3}
}
\end{center}
\end{table*}
Here we present our results for the evolution of a DS, starting from a $3 M_\odot$ star
powered by DM annihilation (defined as $t=0$), building up the stellar mass by accretion, and watching
as the other possible heat sources become important.
    We have considered  a range of DM  masses: 1 GeV,  10 GeV,
     100 GeV, 1 TeV  and 10 TeV, which 
    characterizes the plausible range for 
 WIMP masses.  Since the DM heating scales as $\langle \sigma v \rangle / m_\chi$,
 this range of particle masses also corresponds to a range of annihilation cross sections.
 To account for the uncertainties associated with the 
  accretion rate, the study has used two different accretion
 rates: one from
 Tan \& McKee (2004) and the other from O'Shea \& Norman (2007) (as discussed
 in section \ref{sect:evo} above).
We consider the case of (i) no capture and (ii) ``minimal capture'' (defined as above), with
$\bar\rho_\chi=1.42\times10^{10} {\rm (GeV/cm^3)}$ and
 $\sigma_{sc}=10^{-39} {\rm cm^2}$, chosen so that about  half of the 
luminosity is from DM capture and the other half from fusion for the standard 100 GeV 
case at the ZAMS.  In this paper we will not treat
 the case of maximal capture, where the DM heating would dominate (over all other sources
including fusion) so that the DS would continue to grow as long as it is fed more DM .

We here show the results of our simulations.   First we briefly outline our results and
present our tables and figures.  These are then discussed in detail, case
by case, in the remainder of the results section.  
 
 Tables 1--3, each for a fixed $m_\chi$, present the evolution of some basic parameters of the star throughout the DS phase:
stellar mass $M_\ast$, luminosity $L_\ast$, photospheric radius $R_\ast$, surface temperature 
 $T_{\rm eff}$, the star's baryonic central density $\rho_c$, central temperature 
 $T_c$, the amount of DM inside of the star 
 $M_{\rm DM}$, the DM's central density $\rho_{\rm o,DM}$, as 
 well as the  elapsed time $t$ since the start of the calculation.
 $M_{\rm DM}$ and $\rho_{\rm o,DM}$ give the mass and density of DM either from
 adiabatic contraction (the first 6 rows) or from capture  (the last two rows).
 
 In the tables, the stellar masses were chosen to roughly characterize 
 important transitions for the DS.  The first mass was chosen
 at roughly the time when  a dark star begins to take on the 
 size, temperature, and luminosity which are characteristic of the 
 DS phase, which is around 100 $M_\odot$.  The 
 second mass is given when the DS
 acquires its maximum amount of DM, which varies from 
 case to case.  The third mass is towards 
 the end of the DS adiabatic phase. 
  The fourth row characterizes
  the star's properties just prior to running out of
 adiabatically contracted DM.  The fifth row is given just after the 
 star runs out of adiabatically contracted DM, while the sixth
 gives the star's properties on the ZAMS.  Capture only 
 becomes important  once the adiabatically contracted DM has run out.
 Prior to running out, the stellar evolution is the same with or without DM capture.
 For the case with capture included, the seventh  and eighth row should be used instead of the fifth
 and sixth row. 

Further features of the calculations are shown in Figures 1--4, which
give, respectively,  (1) the evolution of all cases without capture in the
Hertzsprung-Russell diagram, (2) the luminosity as a function of time for the 100 GeV case with
and without capture, (3a) the evolution of the density distribution of  the 
baryonic and dark matter for the 100 GeV case without capture, (3b) the 
distribution of dark matter heating rate for the 100 GeV case with
capture, and (4) the evolution of various quantities  for all cases without
capture.
 
 \begin{figure*}[t]
\centering
\includegraphics[clip=true,width =15cm,height=13cm]{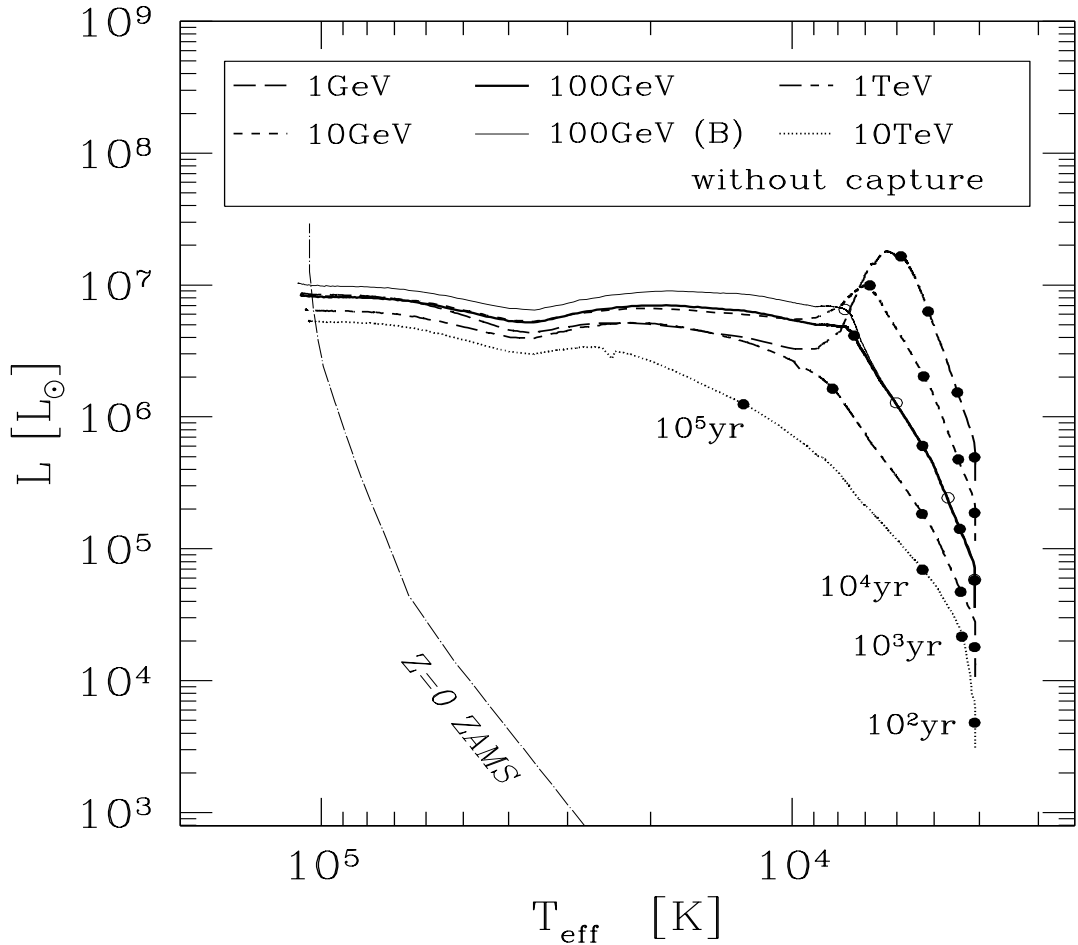}

\caption{ Evolution in the  H-R diagram, for 5 different particle masses 
as indicated in the legend.        
The luminosity is given in solar units and the temperature is in Kelvin.  
The dots indicate a series of time points, which are the same for all 
cases. The open dots distinguish the 100 GeV (B) case from the 100 GeV case.
 All cases are calculated with the accretion rate given by
Tan \& McKee (2004) except the curve labelled 100 GeV (B), which is
calculated with the rate given by O'Shea \& Norman (2007). The metal-free
zero-age main sequence ($Z = 0$ ZAMS) is taken from Schaerer (2002). 
The peak below $10^4$K in the 1 GeV case is due to the overwhelming DM
luminosity in this case.
}
\label{fig:f1}
\end{figure*}

\begin{figure}[t]
\plotone{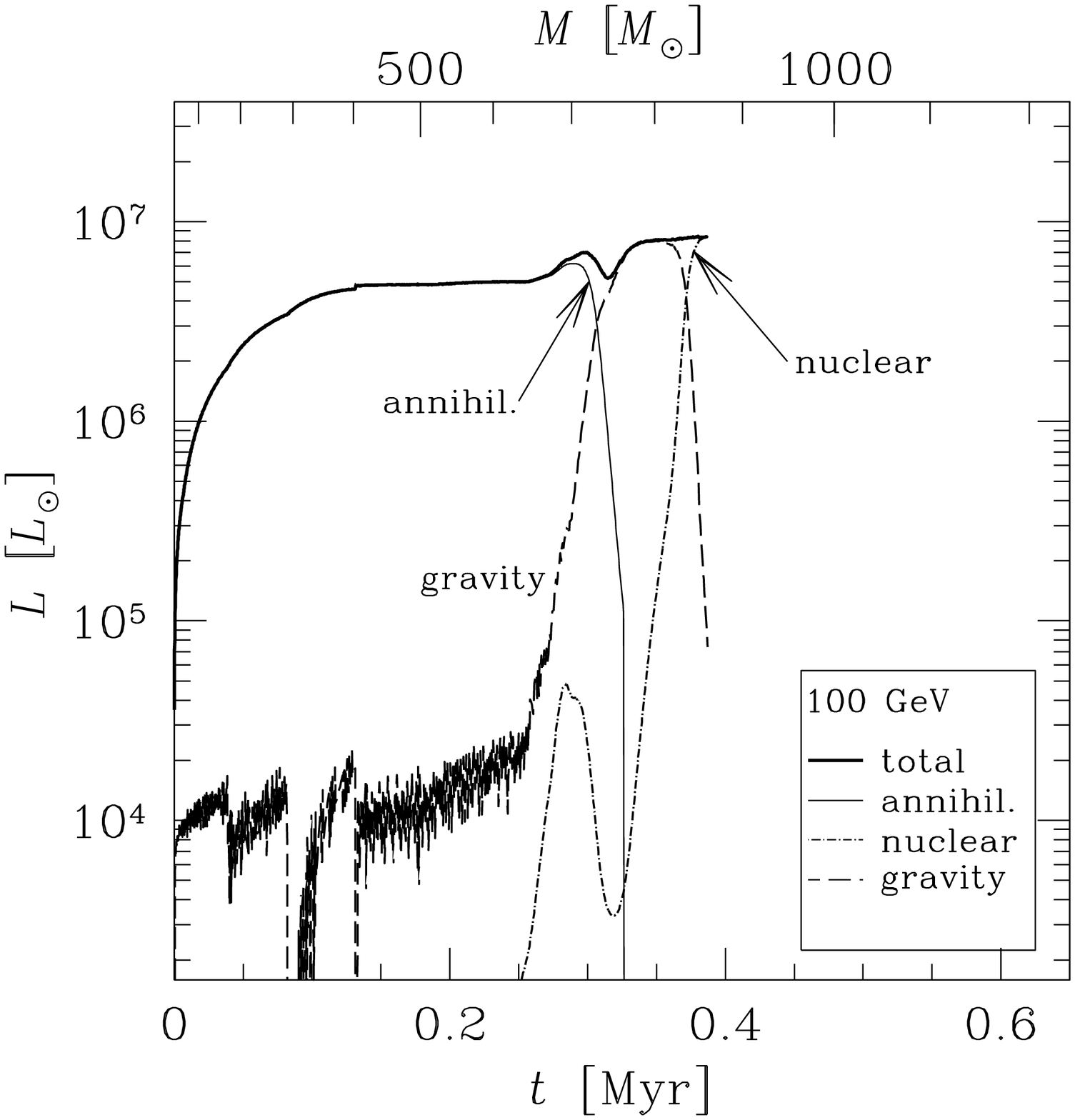}
\plotone{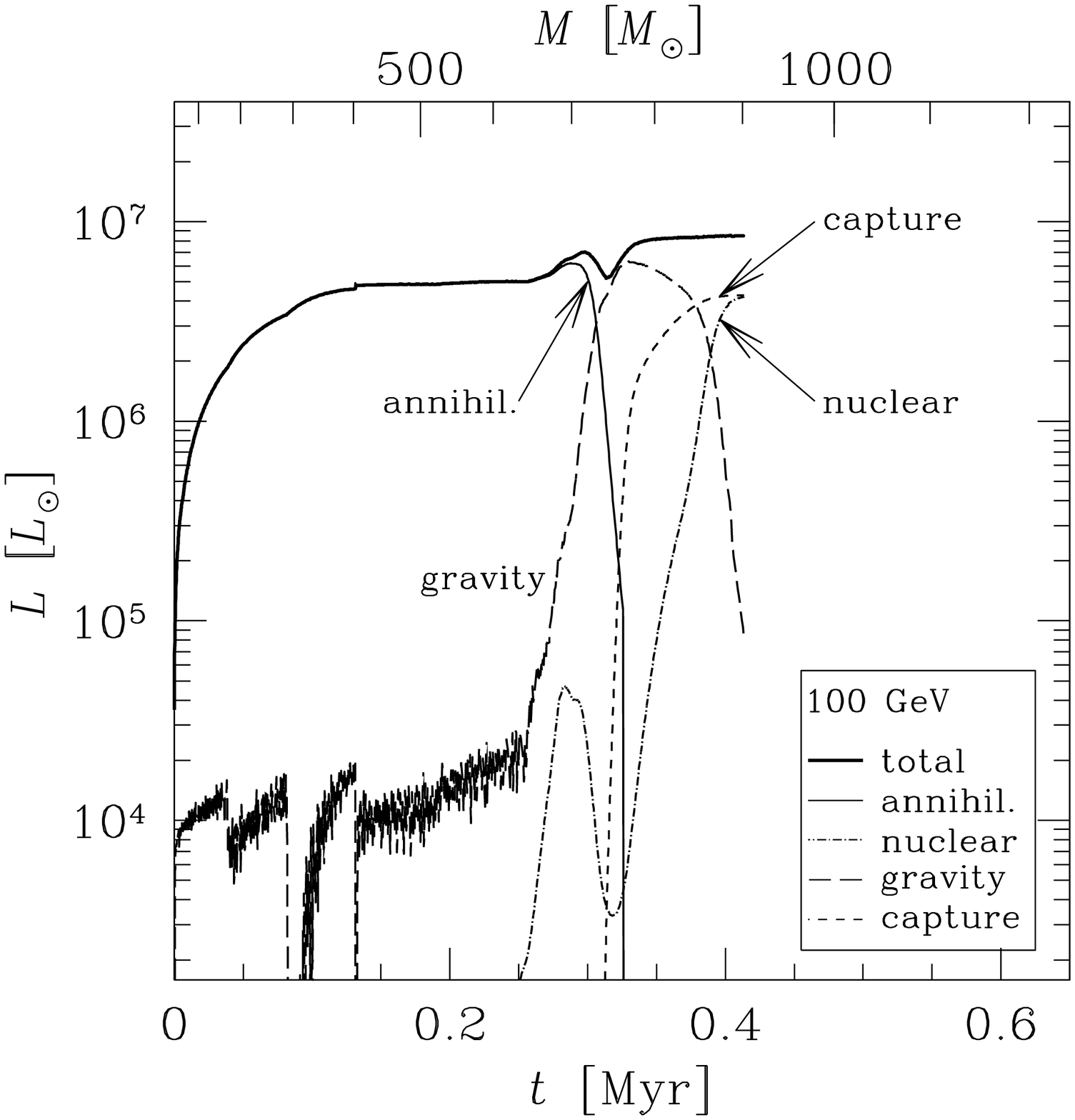}
\caption{Luminosity evolution for the 100 GeV case as a function of time
({\it lower scale}) and stellar mass ({\it upper scale}). The solid (red) 
top curve is the total luminosity.  The lower curves give the partial 
contributions of  different sources of 
energy powering the star  a) ({\it upper frame})
 without capture, and  b) ({\it lower frame}) with `minimal' capture.  In both frames,
 the total luminosity is initially dominated by DM annihilation (the total and annihilation
 curves are indistinguishable until about 0.3 Myr after
 the beginning of the simulation); then  gravity dominates, followed by nuclear fusion.
 In the lower frame, capture becomes important at late times. 
}
\label{fig:f2}
\end{figure}
 
 \begin{figure}[t]
\plotone{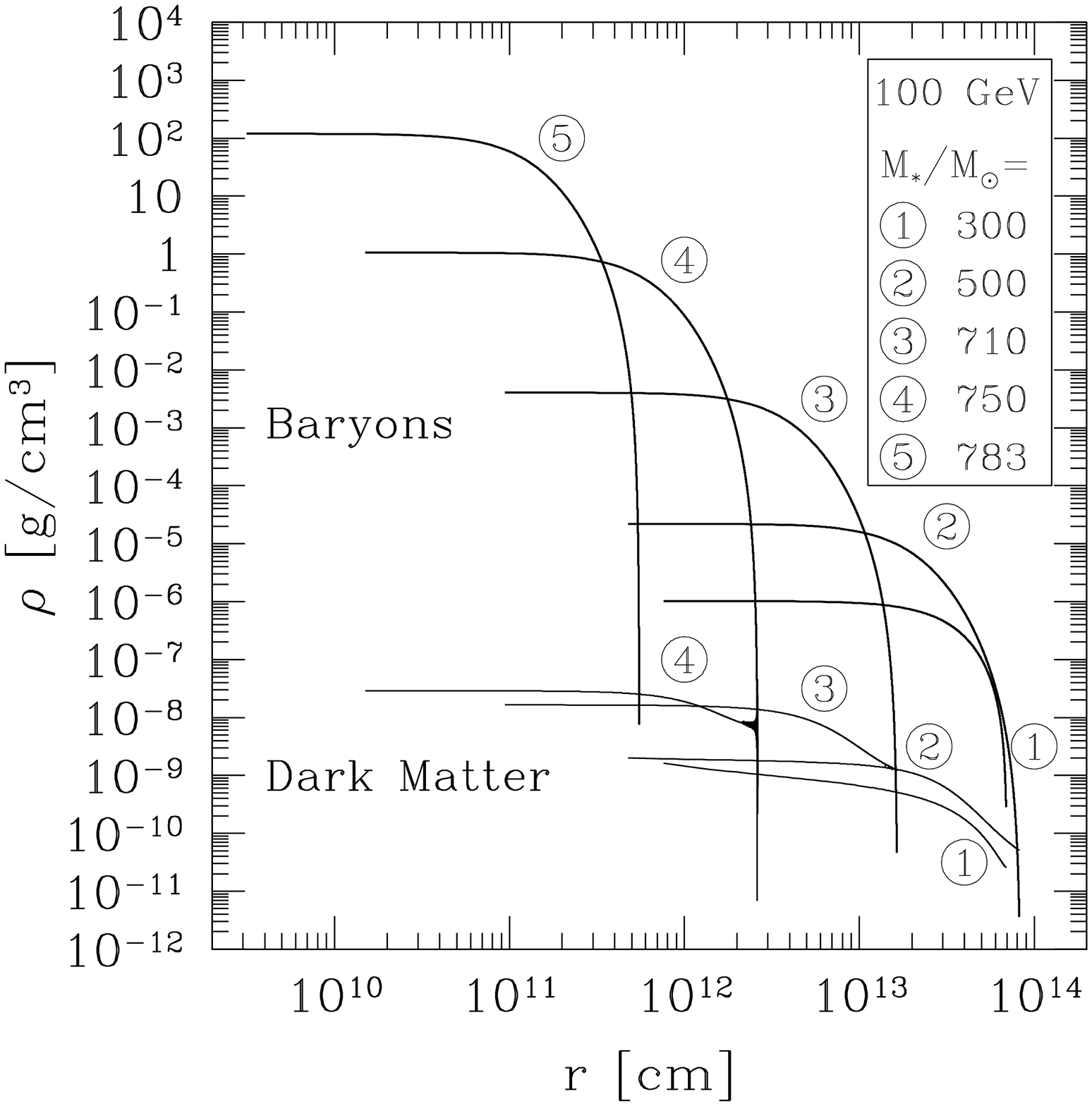}
\plotone{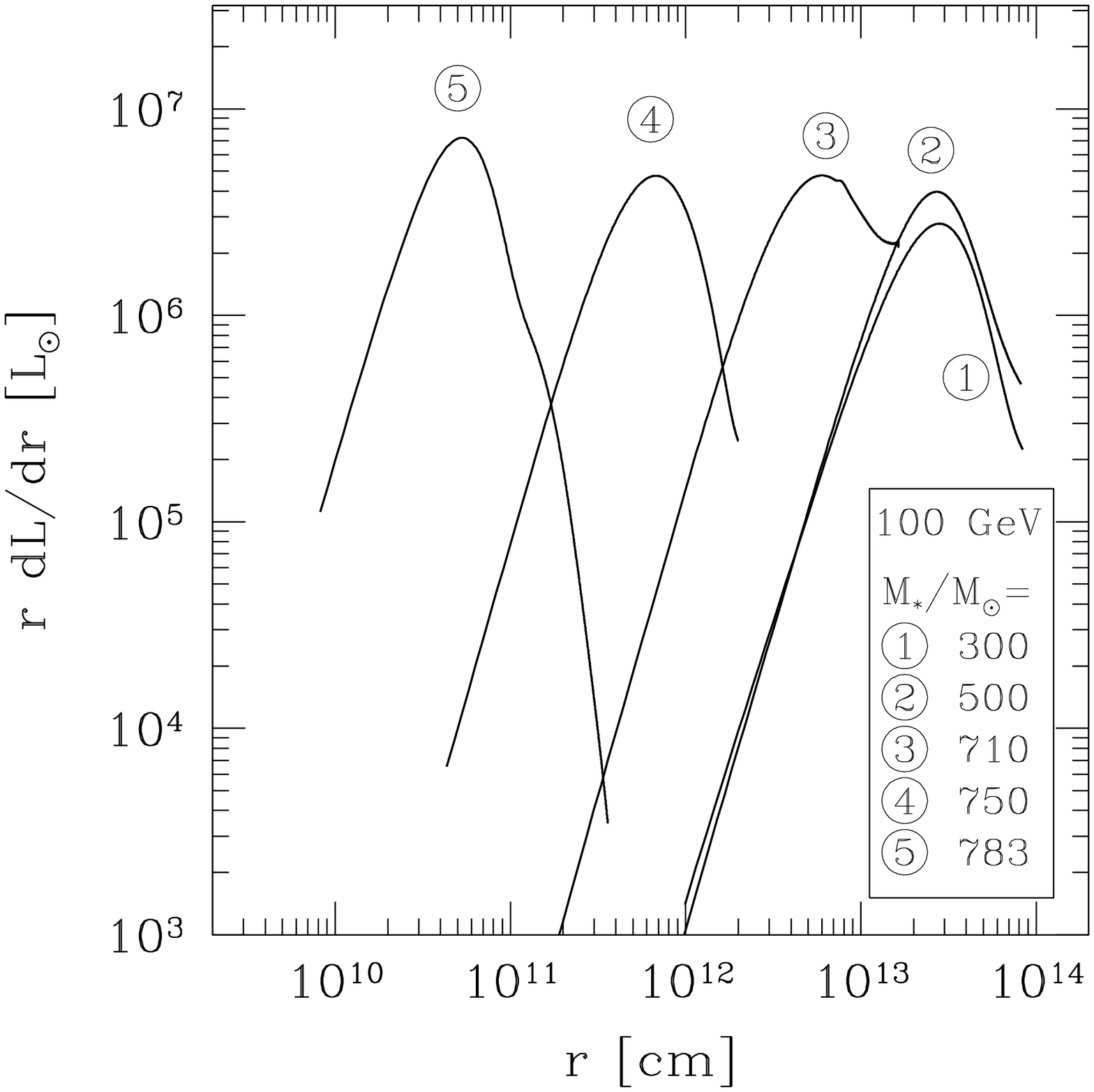}
\caption{ a) ({\it upper frame}): Evolution of a dark star for $m_\chi = 100$ GeV as mass is accreted onto the initial protostellar core of 3 M$_\odot$ (for the case of no capture).  The set of upper
(lower) solid curves correspond to the baryonic (DM) density profile (values given
on left axis) at different stellar masses and times.   
b)({\it Lower frame}): differential luminosity $r dL_{\rm DM}/dr$ as a
 function of $r$ for the masses indicated.        
}
\label{fig:f3}
\end{figure}
 
 \begin{figure*}[t]
\vspace{-.3cm}
\centering
\includegraphics[clip=true,width =15cm,height=13cm]{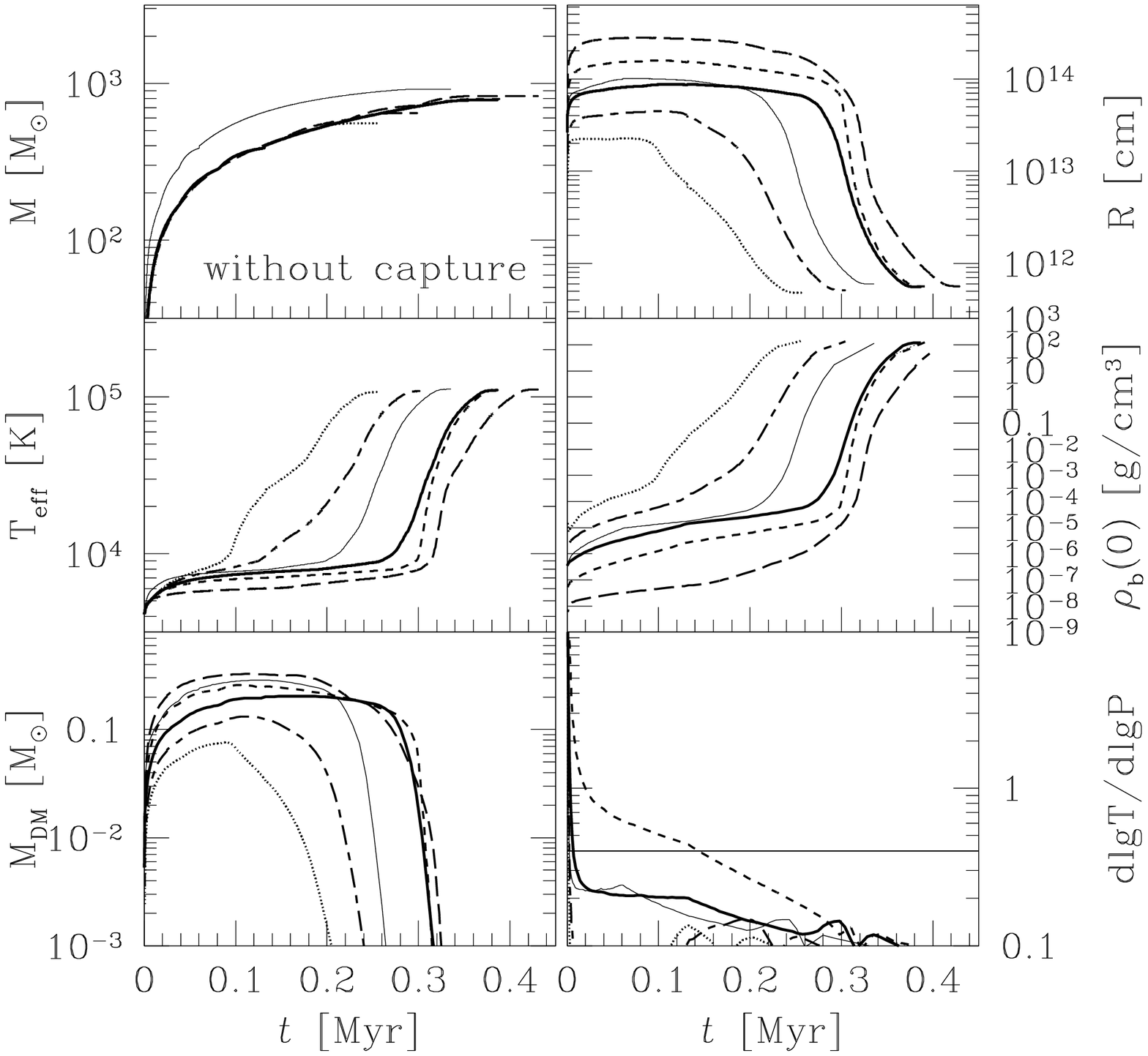}
\caption{
Dark Star evolution (for the case of no capture) illustrating dependence on
DM particle mass  and accretion rate(O'Shea/Norman vs.\ Tan/McKee).
Different curves are the same as explained in the legend of Fig. 1.
The quantities plotted are  a) stellar mass ({\it upper left}), b) total radius 
({\it upper right}), c) surface temperature ({\it middle left}), 
d) central stellar (baryon) density ({\it middle right}), e) amount of DM 
inside of the star ({\it lower left}), and f) radiative gradient at the center
 of the star ({\it lower right}).  In the latter plot,  central regions above 
the horizontal line are unstable to convection; hence this plot illustrates
the onset of a central radiative zone.
}
\label{fig:f4}
\end{figure*}

   \subsection{Canonical Case without Capture}  
   \label{CCwC}
  
   Our canonical case takes $m_\chi=100~{\rm GeV}$,  the 
   accretion rate as given by Tan \& McKee (2004), 
 and $\langle \sigma v \rangle= 3\times10^{-26}~{\rm cm^3 s^{-1}}$ for the 
calculation of the  annihilation rate. 
   
  The initial model (defined to be $t=0$) has a mass of $3~{\rm M_\odot}$  with a 
  surface temperature of $4100~{\rm K}$, a radius of $3\times 10^{13}~{\rm cm}$, a luminosity
  of $4.5\times10^4~{\rm L_\odot}$ and a fully convective structure.  The DS
  stays fully convective for stellar
masses below $50~M_\odot$ at which point a small 
radiative zone appears.  The star makes a transition from convective to
radiative in the $M_\ast=(100-200)~M_\odot$ mass range, with a radiative
zone growing outward from the center, and then becomes fully
radiative (but for a small convective region at the surface) for $M_\ast
> 300~M_\odot$.  The star stays almost fully radiative for the remainder of the
  DS phase.  Above $750~{\rm M_\odot}$, hydrogen and helium burning set in, 
 and a new convective
  zone appears at the center of the star.

 {\it Evolution of Dark Matter Density:} 
The DM densities inside of the star change due to annihilation and the
 evolution of the dark star's baryonic structure. Figure \ref{fig:f3} plots the baryonic
  and DM density profiles for the standard case at several different stellar masses.  As the star's mass 
  grows, the DM and baryonic densities also increase.  
  The  DM profile for the last model, which has
  a stellar mass of $779~{\rm M_\odot}$, is not plotted because at this point 
  all of the adiabatically contracted DM has been used up. 

The adiabatically contracted DM density profile is more or less constant with
radius 
 and then falls off toward the surface of the star.   The DM
 densities for the earlier models show a density increase toward the center of the star due to
 the original NFW profile, e.g. in  the $300~{\rm M_\odot}$ model.  
The slight peak  flattens out at higher stellar masses as can be seen in the
$500~{\rm M_\odot}$ models, since regions with a higher
 DM density annihilate more rapidly than regions with 
a lower DM density.    Once the peak is worn away,
the central DM profile stays flat and ultimately decreases.  

The DM and baryonic densities evolve slowly for much of the DS phase.
In fact  between $100~{\rm M_\odot}$ and $600~{\rm M_\odot}$,  the DM 
central density barely changes (Table 2).
Initially, the DM mass grows (Fig. 4e)  and  reaches a maximum when the
total mass is about 
$450~{\rm M_\odot}$, at which point the star has about 
0.2 M$_\odot$ of DM. 
At this point, the DM heating   becomes insufficient to support the star.  
The star begins to slowly contract and boosts the DM density, but 
 the boost is still modest up to a time of $2 \times 10^5$ years or
until the star reaches $600~{\rm M_\odot}$.
 
Around $600~{\rm M_\odot}$, the star begins to rapidly shrink.  
Note the rapid dominance of gravitational energy at this time
(Fig. \ref{fig:f2}).
The DM densities shoot up, but the total amount 
of DM inside of the star precipitously  drops.  By $750~{\rm M_\odot}$,  
the DM has run out and the star begins to transition
into a standard Pop.~III star.  The star reaches the MS 
at around $780~{\rm M_\odot}$ at which point feedback effects cut off any further accretion.
 
{\it Evolution  of temperature and energy generation:}
The surface properties  of the  DS evolve rapidly between the initial model at  $3~{\rm M_\odot}$ and
  $100~{\rm M_\odot}$.
The luminosity shoots up by a factor of 25, the radius by a factor of 3, and   
the star's surface temperature goes from $4100~{\rm K}$ to $5800~{\rm K}$. 

The star looks radically different from  a metal-free star on the MS.
   Even at $100~{\rm M_\odot}$,
the DS is still very cool with a surface temperature of around
 $5800~{\rm K}$ vs $100,000~{\rm K}$ (for the standard ZAMS) and has a central temperature
of only about a half  a million degrees Kelvin vs 200 million degrees (ZAMS). 
 The star is also much more extended,  $\approx 7 \times 10^{13}~{\rm cm}$
vs $\approx 5 \times 10^{11}~{\rm cm}$ (ZAMS).

Between $100-600~{\rm M_\odot}$, the star's luminosity evolves less rapidly,
 increasing by a factor of 5 to a value of $5\times10^{6}~{\rm L_\odot}$.
The star's surface temperature reaches $7800~{\rm K}$, 
while the radius remains almost constant, passing through a broad
maximum at 
about $8.5\times10^{13}~{\rm cm}$.
The central temperature is still quite cool ($1.8\times10^{6}~{\rm K}$)
 compared to a metal-free 
ZAMS star.  Most of the star is still too cool to even burn deuterium and 
lithium.  
DM continues to power the star, as   is apparent in Figure \ref{fig:f2}.

Above $600~{\rm M_\odot}$, the DS begins to transform into a standard
 metal-free star.  As the DM begins to run out the  surface 
temperature begins to shoot up and  the stellar radius shrinks. Just prior to 
running out of DM,  at $716~{\rm M_\odot}$, the star's surface
reaches $23,000~{\rm K}$, the radius has shrunk to $10^{13}~{\rm cm}$ and
 the central temperature has shot up to 
15 million K. Near the beginning of this rapid contraction the star is
sufficiently hot to burn deuterium rapidly, as can be seen from the 
spike of nuclear heating in Figure \ref{fig:f2} just before 0.3 Myr; this spike lasts about
50,000 years. The burning of 
deuterium, however, remains an insignificant power source for the star.
As the DS runs out of DM
gravity and nuclear burning begin to turn on as energy sources for the star.
Just at the point where all three energy sources (DM heating, gravity, and nuclear burning)
all contribute simultaneously, the 
luminosity increases by about 50 \% and then decreases again; 
one could call this a ``flash" which lasts about 20,000 years and is the ``last gasp"
of adiabatically contracted DM annihilation.  

Near $750~{\rm M_\odot}$ the energy generation in the  star 
becomes dominated by, first,  gravity, and then nuclear burning.
The star's surface temperature is just above 50,000 K (Table 2) and feedback
 effects  have started
to cut off the accretion of baryons.  Eventually, the star begins to
 settle onto the MS and nuclear burning becomes the dominant  
 power source.  The surface temperature has shot up to 109,000 K, 
and finally feedback effects
 have stopped any further accretion.  On the MS, the radius is
  $5.5\times10^{11}~{\rm cm}$ and
 the central temperature has gone up to 280 million K.  At the final cutoff,
 which is our  zero-age main sequence model, the star has a mass of  $779~{\rm M_\odot}$.

The final model on the metal-free MS  compares well 
with  that produced by a full stellar structure code (Schaerer 2002).  
Schaerer's radius 
is slightly larger at $8.4\times10^{11}~{\rm cm}$ vs. $5.5\times 10^{11}
~{\rm cm}$ for the polytrope. 
 The polytrope's surface temperature is slightly too high at 109,000 K vs
 106,000 K
  and     the polytrope's luminosity is too low by a factor of about 2. 
  The difference is  not surprising; we 
have not done the radiative energy transport.
Remarkably, the numbers are quite close despite the polytropic assumption.

\subsection{General Case without Capture}

In addition to the canonical case, we have studied many other parameter choices
as described previously: a variety of WIMP masses (or, equivalently, annihilation cross sections)
and a variety of accretion rates.  In this section we still restrict ourselves to the case of no capture.
In all cases,  we obtain the same essential results.
While one can see in  Figure \ref{fig:f4} 
that there is some scatter of the basic properties of the star, yet we find the following
general results for all cases:
DS evolution occurs  over an extended period of time ranging from $(2-5) \times10^5$ years.
All cases have 
an extended radius $(2\times10^{13}-3\times10^{14}){\rm cm}$, are
cool ($T_{\rm eff}<$10,000 K) for much of the evolution, and have a large 
final mass $(500-1000)~{\rm M_\odot}$.
 All of the  DS  cases studied are physically very distinct from  standard Pop.~III stars. The standard Pop.~III star  is physically more 
compact with a radius of only $\approx 5 \times  10^{11} {\rm cm}$,
and  hotter with a surface temperature of $\approx $100,000 K.  The 
estimated final mass for the standard Pop III.1 star is also
smaller, $\approx 100~{\rm M_\odot}$, because the high surface temperature
leads to radiative feedback effects that prevent accretion beyond that
point (McKee \& Tan 2008).
 
  Figure \ref{fig:f1}   shows the  evolution curves in the H-R diagram 
 for a variety of cases. The figure shows the evolution for various particle masses 
 with the Tan/McKee accretion rate, and compares the 100 GeV case with a 
 calculation using the O'Shea/Norman accretion rate.   

 The DS evolution curves on the HR diagram are qualitatively similar for all of the different cases.
The DS  spends most of its lifetime at low
temperatures below 10,000 K.   Eventually, DM begins to
 run out, the star contracts and heats up, and 
gravity and nuclear burning
  begin to turn on.  Just at the point where all three power sources are important,
  the luminosity "flashes" (as seen in Fig. 2), i.e., increases by a 50 \% for about
  20,000 years and then goes back down.
  Then, DM runs out.  For the lower particle masses, as seen in Fig. 1,
the luminosity drops slightly as 
  gravity becomes the main and dominant power source.  As the surface
   temperature continues to increase, nuclear burning becomes important  
   when $T_{\rm eff}\approx 30,000$ K. Above this temperature (see Fig. 1), the stellar luminosity  
again increases as the star approaches the ZAMS.  Above 50,000 K,
 feedback effects turn on and begin to stop the accretion of baryons.  Once the star reaches 100,000 K, 
 accretion completely stops, and  the star goes onto the ZAMS.

While the basic DS evolution is similar for all cases, we now discuss the variation
due to different parameters.
The evolution curves differ depending upon the DM particle mass.
For larger $m_\chi$,
the stellar  mass and luminosity decrease such that the evolution 
curve on the H-R diagram for the 1 TeV case is beneath  that for the 100 GeV case. The reduced
luminosity is a consequence of the reduced energy production at the
higher particle masses, since DM heating is inversely proportional
to $m_\chi$  (see eq. (\ref{eq:e1})).  Then at higher particle masses a smaller radius is needed to 
balance the dark-matter energy production and the radiated luminosity.

In the Tables and in Figure 3, one can see the density profiles of the various components in the DS.
The DM densities also depend upon $m_\chi$.
As shown in the Tables, the average DM density in the star is an increasing function of $m_\chi$.
For example, at the time the DS reaches 100 $M_\odot$, the central DM densities vary between
 $2 \times 10^{-8}~{\rm g/cm^3}$ for $m_\chi=10~{\rm TeV}$ and
$ 5 \times 10^{-11}~{\rm g/cm^3}$ for $m_\chi=1~{\rm GeV}$.
These densities are many orders of magnitude lower than the baryon densities in the DS,
as can be seen in Figure 3.  Here one can see ``the power of darkness:"  Despite the low
DM densities in the star, DM annihilation
is an incredibly efficient energy source since nearly all of the particle mass is converted to heat
for the star.  We also wish to point out that both baryon and DM densities in the DS
are far lower than in metal free main sequence stars of similar stellar mass; ZAMS stars
have densities $> 1$ gm/cm$^3$.

One can also see the dependence of the DS evolution upon accretion rate.
Our canonical case, obtained using the accretion rate from Tan \& McKee (2004),
can be contrasted with that using  the O'Shea/Norman accretion rate.
For 100 GeV particles, the DS evolution is identical for the two accretion rates until
$T_{\rm eff} \sim 7,000$ K (10$^5$ yr).  After that the luminosity of the
O'Shea/Norman case  shoots higher  as a result of 
the higher accretion rate.  Consequently the DS with the
   O'Shea/Norman accretion rate has a shorter lifetime than the DS with the
   Tan/McKee accretion rate
   (350 vs. 400 thousand years); a larger final mass 
 ($916 {\rm M_\odot}$ vs. $780 {\rm M_\odot}$) and higher luminosity.

 The 1 GeV case displays some interesting behavior:
one can see a spike in the evolutionary track on the H-R diagram.
The DS reaches a maximum luminosity of $2\times10^{7}~{\rm L_\odot}$ and
 then subsequently drops by a factor of nearly ten, 
 crossing beneath the 100 GeV case.  Subsequently 
 the 1 GeV curve then again crosses the 100 
 GeV case before settling onto the MS.
The large early luminosity is due to extreme effectiveness
 of DM heating for the 1 GeV case; DM heating is inversely proportional 
to $m_\chi$ (eq. \ref{eq:e1}).

 The stellar structure of all of the
 cases studied is similar to our canonical
case, but the transition from a convective to  a 
  radiative structure  happens at different times, depending upon the 
  DM particle mass.  The lower right panel in Fig. 4 plots 
  the temperature gradient in the innermost few zones at the center of the DS,
  to show the onset of radiative energy transfer at the center of the star.
  Models above the line have convective cores while models below have
  radiative cores.  As we increase the DM mass, the onset of radiative transport
  happens earlier, at lower stellar masses. 
   For our canonical 100 GeV case, a small radiative zone appears when the
  mass of the DS reaches 50$M_\odot$.  The $m_\chi=1~{\rm GeV}$ case, however,
   does not begin to form a radiative zone until 
  the mass is nearly $400~{\rm M_\odot}$. The larger radii and cooler
internal temperatures, compared with the canonical 100 GeV case, result in higher
interior opacities, favoring convection.

  The timing of the transition from convective to radiative can be understood from the stellar structure.   For $m_\chi$ greater than the canonical case, the DM heating is less effective
  since it is  inversely proportional to $m_\chi$.  The star compensates by
becoming more compact, thereby increasing the DM density and the internal temperature.
In the relevant temperature range, the opacity scales as $\rho T^{-3.5}$, and
$\rho$ scales as $T^3$. The smaller radius actually results in a 
decrease in luminosity as well, so the radiative gradient
\begin{equation}
\frac{{\rm d~log} T}{{\rm d~log} P} = \frac{3 \kappa L P}{16 \pi G ac T^4 M_r}
\end{equation}
is reduced, tending to suppress convection relatively early.
  Conversely if $m_\chi$ is less than $100{\rm GeV}$, 
  the star must be more puffy, which increases the radiative gradient.
  Thus the transition will occur at a later stage. 

 The cases studied have some differences in terms of lifetime, luminosity, and
  final mass.  At the time the DS has grown to $100~{\rm M_\odot}$,  
 the 1 GeV case is an order of magnitude brighter than the 100 GeV case.
  Similarly, the 100 GeV case is about an 
 order of magnitude brighter than the 10 TeV case. The DS lifetime grows shorter
  for increasing $m_\chi$. The lifetime for 
 the 1 GeV case is $4.3 \times 10^5$  years while the lifetime for the
 10 TeV case is only $2.6 \times 10^5$ years. The reduced amount of 
fuel available in the
10 TeV case more than compensates for the lower luminosity in determining the
 lifetime.  Also, the 1 
 GeV case has a higher  final mass  than the 10 TeV case ($820~{\rm M_\odot}$
 vs $552~{\rm M_\odot}$); clearly a longer lifetime allows more mass to accrete.

 \subsection{ DM Evolution-Adiabatic Contraction Only}
 
 Here we discuss the evolution of the DS once its DM fuel begins to run out. 
 In this section, we do not consider capture; in the next section we will add the effects
 of capture as a mechanism to refuel the DM inside the star.

Initially, as baryons accrete onto 
the DS, adiabatic contraction pulls  DM into the
 star at a sufficiently high rate to support the star, 
but eventually,  adiabatic contraction fails to pull in 
enough DM at high enough densities to support it.        
 
Adiabatic contraction eventually fails for two main reasons.  
DM must be brought inside the star, first, to replace the DM which has 
annihilated away and, second, to support an ever-increasing stellar 
luminosity as the stellar mass increases due to accretion.  In fact, an
accretion rate  that decreases as a function of time (Tan \& McKee 2004) 
only compounds the problem, leading to a relatively shorter life span since
 even less DM is brought into the star.  With an insuffient amount of DM 
at high enough densities, the star must contract to increase the DM 
heating rate to match the stellar luminosity. The contraction leads to
higher surface temperatures such that feedback radiative effects 
eventually shut off accretion altogether.

As the star contracts, a large fraction of DM previously inside of the 
star ends up outside of the star. The star's contraction amounts to 
at least two orders of magnitude in radius, 
leaving a 
large fraction of the DM outside the star. Thus DM annihilation accounts for only a 
fraction of the DM ``lost" from the star.  For instance in the 100 GeV case,
 only a fifth of the DM is lost due to annihilation; the rest is ``left outside".

The ``left outside" effect can be understood in a simple way.  
DM heating is mainly a volume effect as can be
seen in Figure \ref{fig:f3}.  Due to adiabatic contraction, 
$\rho_{\rm DM}$ does not scale perfectly with  
baryon density  (eq. \ref{eq:AC}).  For instance, 
the total amount of DM inside of star of fixed mass (even without DM
annihilation) decreases as the radius of the  star shrinks.  

Without a way to replenish the DM such as with capture,  the DS will eventually run out of DM 
due to the combination of the ``left outside" effect and DM annihilation.

\subsection{DS evolution with Capture included}

The DM inside the DS can be replenished by capture.  DM particles bound
to the larger halo pass through the star, and some of them can be captured by losing
energy in WIMP/nucleon scattering interactions.
 We remind the reader that we have  defined our case of 'minimal capture' to have
  background DM density $\bar\rho_\chi =1.42\times10^{10} {\rm (GeV/cm^3)}$ and scattering cross section
 $\sigma_{sc}=10^{-39} {\rm cm^2}$ chosen so that about  half of the 
luminosity is from DM capture and the other half from fusion for the standard 100 GeV 
case at the ZAMS.  

  With capture, the stellar evolution is the same as in the 
cases previously discussed without capture throughout most of the history of the DS.
The contribution to the heating from capture  $L_{\rm cap}$ only becomes 
  important once DM from adiabatic contraction runs out.     Once this
happens,  gravity still initially dominates, but  for the 'minimal capture' case we have considered,
  $L_{\rm cap}$ is  more important than $L_{\rm nuc}$ until the  star eventually 
  reaches a new ``dark"  main sequence, where the two contributions become
equal.
  During this phase
  the star is more extended and cooler than 
  without capture, which allows for more baryons to accrete onto the star
 once the first dark star phase driven by adiabatic contraction ends. 
However for the standard minimal case the final mass with capture is at most 1\%
 larger than that without.
  
 In a future paper we will consider the case of much higher capture rates than
 the minimal ones we have studied here. The capture rate depends on the ambient
 DM density in which the DS is immersed and on the WIMP/nucleus scattering cross section.  High ambient density  could easily arise; in fact the
 adiabatic contraction arguments applied to the DS from the beginning indicate that
 this is quite likely.  On the other hand, the scattering cross section may be significantly lower
 than the current experimental bound adopted in our minimal capture case.
 In any case, with a sufficiently high capture rate, say two or more orders of
magnitude higher than the minimal value considered here,  the star can stay DM powered and sufficiently
 cool such that baryons can in principle continue to accrete onto the star
 indefinitely, or at least until the star is disrupted. This latter case will be explored in
  a separate paper where it will be shown that the dark star could 
easily end up with a  mass on  the order of several tens of thousands of solar masses 
  and a lifetime of least tens of millions of years.

\section{Discussion}
We have followed the growth of  equilibrium protostellar Dark Stars (DS), 
powered by Dark Matter (DM) annihilation,
 up to the point when the DS descends onto the main sequence.  Nuclear burning,
gravitational contraction, and DM capture are also considered as energy
sources. During the phase
when dark-matter heating dominates, the objects have sizes of a few AU and
central $T\approx 10^5-10^6$ K.                                 
Sufficient DM  is brought into
the star by contraction from the DM   halo to result in a 
DS phase of at least several hundred thousand years (without DM capture).  Because of the 
relatively low $T_{\rm eff}$  (4000--10,000 K), feedback 
mechanisms for shutting off
accretion of baryons, such as the formation of HII regions or the dissociation
of infalling H$_2$ by Lyman-Werner photons, are not effective until DM begins to run out.
The implication is that main-sequence stars of Pop.~III are very massive.
Regardless of  uncertain parameters such as the DM particle
mass, the accretion rate, and scattering, DS  are cool, massive,
 puffy and extended. The final masses lie in the range 500--1000 M$_\odot$,
very weakly dependent on particle masses, which were assumed to vary
over a factor of $10^4$.

One may ask how long the dark stars live.  If there is no capture, they live until the
 DM they are able to pull in via adiabatic contraction runs out; the numerical
results show lifetimes in the range $3 \times 10^5$ to $5 \times 10^5$ yr.
If there is capture, they can continue to exist as long as they
 reside in a medium with a high enough density of dark matter to provide their
entire energy by scattering, capture, and annihilation.

In addition, the properties of the DM halo affect  the nature of the DS.  For instance, a larger concentration parameter allows for more DM to be pulled in 
by adiabatic contraction. The additional DM prolongs the lifetime of the DS phase  and increases the DS final mass.
The canonical case with a concentration parameter c $=3.5$ rather than x $=2$ has a final mass of 870 $\rm M_\odot$ vs. 779 $\rm M_\odot$
and has a lifetime of 442,000 yrs vs. 387,000 yrs. Furthermore, the concentration parameter of each DS halo will be different, depending upon the history of each halo.
Thus the Initial Mass Function (IMF)  of the ZAMS arising from the growth of dark stars will be sensitive to the halo parameters. This effect will be studied thoroughly in a separate publication. 

DS shine with a few $10^6 L_\odot$  and one might hope to detect them,
particularly those that live to the lowest redshift ($z=0$ in the most optimistic case).
 One may hope that the ones that form most recently
are detectable by JWST or TMT and differentiable from the standard
metal-free Pop.~III objects.
DS are also predicted to have atomic hydrogen lines originating in
the warmer photospheres,  and H$_2$             
lines arising  from the infalling material, which is still relatively cool.

The final fate of DS once the DM runs out is also different from that of standard Pop III stars,
again leading to different observable signatures.   
Standard Pop III stars are thought to be $\sim 100-200 M_\odot$, whereas
DS lead to far more massive MS stars.
Heger \& Woosley (2002)    showed that for $140
M_\odot < M < 260 M_\odot$, pair instability supernovae lead to
odd-even effects in the nuclei produced, an effect that is not observed.
 Thus if Pop III.1 stars are really in this mass
range one would have to constrain their abundance. Instead, we find
 far more massive Dark Stars which would  eventually become heavy main sequence (ZAMS)
 stars of 500-1000 $M_\odot$.
 Heger \& Woosley find that these stars after relatively short lifetimes collapse to black holes.
Or, if they rotate very rapidly, Ohkubo et al (2006) argue that they could become supernovae, leaving
behind perhaps half their mass as black holes. In this case the 
presumed very bright supernova could possibly be observable.
In either case, the elemental abundances arising from pair instability SN of standard Pop III
(less massive) stars are avoided so that one might avoid the constraint
on their numbers.  Other constraints on DS will arise from cosmological considerations.
A first study of their effects (and those of the resultant MS stars) on reionization have been done by Schleicher et al. (2008a,2008b), and further work in this direction is warranted.

The resultant black holes from DS could be very important.
DS would make plausible precursors of the $10^9 M_\odot$
black holes observed at $z=6$ (Li et al. 2007; Pelopessy et al. 2007); of
Intermediate Mass Black Holes; of black holes at the centers of galaxies; and
of the black holes recently inferred as an explanation of the extragalactic radio excess seen by  the ARCADE experiment (Seiffert et al. 2009).  However, see
Alvarez et al. (2008) who present  caveats regarding the growth of early black holes.
In addition, the black hole remnants from DS could play a role in
  high-redshift gamma ray bursts thought to take
place due to accretion onto early black holes (we thank G. Kanbach for 
making us aware of this possibility).

In the presence of prolonged capture, the DS could end up even larger, possibly leading
to supermassive stars that consume all the baryons in the original halo,
i.e.,  possibly up to $10^5 M_\odot$ stars. The role of
angular momentum during the accretion must be accounted for properly and will affect
the final result; in fact, even for the $1000 M_\odot$ case studied in this paper,
we intend to examine the possible effects of angular momentum.  This possibility will be further investigated
in a future publication (in preparation), both from the point of view of the stellar structure, as well
as the consequences for the universe. 

We acknowledge support from: the DOE and MCTP via the Univ.\ of
Michigan (K.F.); NSF grant AST-0507117 and GAANN (D.S.); NSF grant
PHY-0456825 (P.G.).  K.F. and D.S. are extremely grateful to
Chris McKee and Pierre Salati for their encouragement of this line of
research, and to A. Aguirre, L. Bildsten, R. Bouwens, N. Gnedin, G.Kanbach, A. Kravtsov, N. Murray, J. Primack,  
C. Savage, J. Sellwood, J. Tan, and N. Yoshida for helpful discussions.

\end{document}